# Space-filling, multifractal, localized thermal spikes in Si, Ge and ZnO

Shoaib Ahmad[1,2,a], M. Sabtain Abbas[1], M. Yousuf[1,3], Sumera Javeed[4], Sumaira Zeeshan[4] and Kashif Yaqub[4]

[1]CASP, Church Road, Government College University (GCU), Lahore 54000, Pakistan

[2]Abdus Salam National Center for Physics, QAU Campus, Islamabad 44000, Pakistan

[3]Dept. of Physics, University of Engineering and Technology, Lahore 54000, Pakistan

[4]PINSTECH, P O Nilore, Islamabad 44000, Pakistan

[a]Corresponding Author: sahmad.ncp@gmail.com

**ABSTRACT**

The mechanism responsible for the emission of clusters from heavy ion irradiated solids is proposed to be thermal spikes. Collision cascade-based theories describe atomic sputtering but cannot explain the consistently observed experimental evidence for significant cluster emission. Statistical thermodynamic arguments for thermal spikes are employed here for qualitative and quantitative estimation of the thermal spike-induced cluster emission from Si, Ge and ZnO. The evolving cascades and spikes in elemental and molecular semiconducting solids are shown to have fractal characteristics. Power law potential is used to calculate the fractal dimension. With the loss of recoiling particles' energy the successive branching ratios get smaller. The fractal dimension is shown to be dependent upon the exponent of the power law interatomic potential $D = \frac{1}{2m}$. Each irradiating ion has the probability of initiating a space-filling, multifractal thermal spike that may sublime a localized region near the surface by emitting clusters in relative ratios that depend upon the energies of formation of respective surface vacancies.

**Keywords:** Thermal spikes, collision cascades, fractal dimensions, space-filling multifractal

# 1. Introduction

Clusters of negative atoms have long been observed, along with monatomic species, being emitted from cesium ion irradiated solids in the Source of Negative Ions by Cesium Sputtering-SNICS. The device was developed to deliver anions to the positively charged high voltage terminals in Tandem accelerators [1]. The present study has been conducted on a SNICS source with elemental (Si and Ge) and molecular (ZnO) targets. Mass spectrometry of the sputtered species is performed as a function of $Cs^+$ energy-$E(Cs^+)$. This study is the extension of the similar work on fullerite, graphite and the single and multi-shelled carbon nanotubes [2-4]. The objective is to verify and quantify cluster emission and to identify the physical mechanism. Linear collision cascade–based sputtering theories [5-7] successfully explain the sputtering of atoms from elemental metals. The atomic collision cascades that evolve in the sub-surface layers under the irradiated surface are assumed to be the source for the sputtering of atomic constituents. Energy of direct recoils is shared in the cascades of binary collisions. Sputtering yield, therefore, counts the number of recoiling target atoms leaving the outer surface per incident ion. Clusters of atoms are neither considered nor expected to be sputtered from the cascades. The sputtering yields of metallic targets irradiated with low to medium mass ions could be described by these linear energy deposition-based binary collision cascades [5-7]. On the other hand, the nonlinear energy deposition by the irradiation of heavy ions, charged clusters and multiply charged ions in solids were suggested to initiate localized thermal spikes. Spikes were used to explain cluster emissions, the much higher sputtering yields and other associated radiation damage. Theoretical models to explain the spike phenomenon varied considerably in their assumptions that ranged from cascade overlap to the crater formation [8-13]. These models had specific applicability to the irradiations of semiconductors and insulators.

An ion-induced binary collision cascade in a solid is a typical example of Mandelbrot's fractals [14]. The fork-like branching pattern of the colliding and recoiling atoms is self-similar and scale invariant. However, near the end of the cascade, the collision mean free path approaches atomic spacing, energy is dissipated in randomized atomic motion that has the potential of generating a local hot spot. A thermal spike may ensue if the end-of-the-cascade terminations overlap. In the context of the evolving and emerging fractal description of the transition from collision cascades to thermal spikes, we invoke multifractal character where the fractal dimension varies depending upon the nature of the ion-target and target-target combinations. In this communication, we elaborate the space-filling, multifractal characteristics of the kinetic phase transition of cascades into thermal spike by utilizing mass spectrometric data of $Cs^+$ irradiated elemental semiconductors Si and Ge and molecular semiconductor solid ZnO as a function of the energy of $Cs^+$ ions.

Similar examples of fractals have emerged from radiation damage studies involving cracks and fractures in addition to sputtering by cascades and spikes. The earliest investigations of the fractal character of fracture were in metallic and ceramic solids [15-17]. Cherapanov's review [18] on fractal fracture mechanics provides a broad overview of the rapid progress in the field within the two decades after Mandelbrot introduced the geometry of the evolving symmetry of fractals. The fracture solid mechanics describes a deformed solid as a thermodynamically open, non-equilibrium system where the self-organization of the dissipative structures (cracks, pores and cascades) occurs. In this communication, the spreading of the cascade is visualized as self-similar at different scales. Linear energy deposition by cascades of binary collisions were successfully employed in the early radiation damage theories to estimate the sputtering yields, the number of vacancies and the nature of the radiation damage. Sputtering yields were shown to be linearly proportional to the energy deposition function. Nonlinear energy depositions produced by the irradiations of slow, heavily charged ions were shown to produce multiple types of damages, for example, highly charged $Th^{70+}$ and $Au^{69+}$ were shown to cause the emission of the negatively and positively charged carbon clusters $C_x^+$ and $C_x^-$ from HOPG and $C_{60}$, where x≤20 [19]. Similar results were reported by irradiations with Au cluster ions [20] and $Cs^+$ irradiations of carbon's nanostructures and graphite [2-4].

We have chosen Si and Ge for the present study because the spike-induced radiation effects and the damage on Si and Ge digital devices are of considerable interest and subject of active investigations [21-22]. Local heating and melting of the ion deposited energy lead to amorphous regions in electronic circuits. Higher doses can produce swelling; void growth and anomalous structure formation within Si and Ge. Molecular dynamics (MD) simulations of the nature of defects generation in metals, Si and Ge had a long and fruitful history. MD studies of cluster emission from irradiated Cu and Ag demonstrated the nature of the spikes that were generated [23,24]. Another example of the attempts to understand the nature of cascades in molecular solids is where low energy $Xe^+$ irradiation of molecular benzene crystals was simulated by MD [25]. The study pointed to the thermal origin of sputtering. Extensive MD simulation studies of the production of defects in Si and Ge by thermal spikes has provided better understanding of the mechanism and its effects [26]. Local heating by thermal spikes lead to the melting of the surrounding regions; often well below the outer surface. This introduces amorphous spots within the electronic circuitry. Crater-like surface defects are also produced on irradiated solids. A MD simulation of the energy regimes that lead to crater formation deals with the transition from binary collision cascade to thermal spikes has been illustrated in 0.4 to 100 keV $Xe^+$ irradiations of Au [27]. The authors of the study worked out the dependence of crater size on the irradiating ion energy and its relation with the lifetime of the heat spike. The duration of thermal spikes in various studies on a range of solids ranged between 0.05 to 2 ps [28-31]. The ion-solid combinations and the material properties (specific heat, density, thermal conductivity) along with the radial dimensions of the core of the thermal spike determine the spike duration.

The extensive range of experimental and simulation studies [19-31] have demonstrated the generation of thermal spikes. We have recently worked out a statistical thermodynamic model where the probability of creation of thermally activated mono- and multi-vacancies on the surface of single-walled carbon nanotubes-SWCNTs, is given as $p_x \propto (\exp(E_{xv}/kT) + 1)^{-1}$, where $E_{xv}$ is the energy of formation of $x$-member vacancy and $T$ is the sublimation temperature [4]. The model was applied to explain the mass spectrometric results of the emission of carbon clusters $C_x^-$; $x \geq 1$ from $Cs^+$ irradiations of SWCNTs. Here we present the mass spectrometric results of cluster emission from Si and Ge irradiated by $Cs^+$ in the energy range 0.5 to 5.0 keV. These results are then compared and contrasted with the sputtering of molecular solid ZnO irradiated with $Cs^+$ of the same energy range. It is shown here, by comparing experimental results with the calculated rates of ZnO dissociation into Zn and O that one can establish the existence of thermal spikes in ZnO as well as Si and Ge. We then describe the transition from binary collision cascades into the locally heated zones of thermal spike by using analogy with fractals. The kinetic phase transition from cascades into spikes is described as space-filling, multifractals. The dynamics of the transition is reflected in the evolving fractal dimension $D$.

## 2. Experimental

We employed Source of Negative Ions by Cesium Sputtering-SNICS [1] equipped with a momentum analyzer as the experimental set-up for ejecting the target material at a desired rate by sputtering while the parameters of the irradiating ion $Cs^+$ can be accurately controlled. The source was operated by varying energy of the $Cs^+$ ions $E(Cs^+)$ between 0.5 and 5.0 keV. The Cu bullets containing Si, Ge and the sintered ZnO pellet were used as targets for NEC's SNICS II negative ion source mounted on the 2MV Pelletron at Government College University (GCU), Lahore. ZnO pellet was prepared from ZnO powder 99.99% pure ZnO powder was pressed and then sintered. The ZnO pellet was then inserted into Cu bullet. The sputtered negative ions, the anions of $Si_x^-$, $Ge_x^-$; $x \geq 1$, consist of atomic and cluster species. From the irradiated ZnO the anions $O^-$, $O_2^-$, $ZnO^-$ and $ZnO_2^-$ were extracted. All anions were extracted from the source at constant beam energy while the target bias i.e. $E(Cs^+)$ was adjusted between 0.5 and 5.0 kV. The combination of the extraction voltage and acceleration voltages was maintained at a constant 20 kV. This ensured that the sputtered ions are extracted with the same energy even though these were subjected to varying $Cs^+$ energies $E(Cs^+)$. The high electron affinities of the sputtered species facilitate their detection. The species are sputtered predominantly in neutral state and detected subsequently as anions after passing through the neutral $Cs^°$ coated target surface and the electron cloud in the ionizer. A 30° bending magnet analyzed the anions. Detection of negative anions has its advantages over that of the positively charged cations. Hot plasmas are required for the production of positive charges to remove at least one electron from the respective species. Larger clusters are likely to fragment in collisions with hot electrons and other charged ions. SNICS operates at low temperature (∼100°C) at which neutral $Cs^o$ ionizes. It is a preferable, low temperature collision chamber to study the sputtered neutral species by conversion into anions with the attachment of electrons after ejection from the irradiated surface.

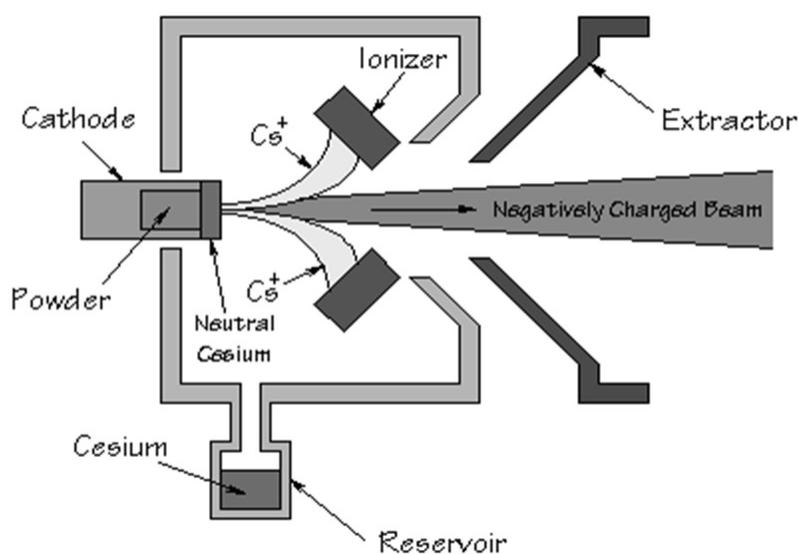

Fig. 1. Schematic drawing of the SNICS source.

The design of SNICS is to optimize sputtering of the targets with $Cs^+$ at 4-5 keV energies. At these energies, a well defined beam spot of around 1 mm diameter is produced at the center of the target. This ensures maximum yield of the sputtered, negatively charged species that are mass selected and sent to the Tandem accelerator. The accelerating and focusing voltages are optimized around this central design feature of SNICS. However, at lower $Cs^+$ energies, one gets a broad $Cs^+$ beam that bombards the whole surface of the target and has a diameter of ∼ 4 - 5 mm. This allows larger surface area of the target to be irradiated at energies ≤ 2 - 3 keV. It serves two purposes; ion density per unit surface area remains low and secondly the whole target is equally irradiated. One also avoids excessive, irradiated Cs build-up near the surface of the target. Considering the SNICS's basic design features, the low energy operation is not an efficient way of using it as high current source for Pelletron accelerator, but from the point of view of the controlled surface damage studies that require variable energy ion-induced sputtering of multi-atomic clusters, the low $Cs^+$ energy irradiations are preferred. The experiments reported in this study have been conducted by operating SNICS where the cluster emissions can be investigated as a function of $E(Cs^+)$.

## 3. Experimental Results

### 3.1 Sputtering of Si, Ge and ZnO in SNICS

Figures 2 and 3 show the normalized number densities of the sputtered cluster anions of Si and Ge species as $Si_x^-$ and $Ge_x^-$ where x≥1. The data has been obtained from the mass spectra of the sputtered species by varying the energy of the $Cs^+$ ion beam between 0.5 and 5.0 keV. The low energy irradiations ensure the smaller ionic ranges and the sub-surface collision cascades. The $Cs^+$ ion ranges in this energy range are of the order of few nm. The normalized yields of the four sputtered species is plotted as a function of the $E(Cs^+)$ in Fig. 2. At $E(Cs^+)$ = 0.5 keV, the diatomic $Si_2$ as three times higher yield as compared with the monatomic $Si_1$. The emission at the low ion energies with low beam current (<0.01 mA) and spread over the whole target surface (>12 $mm^2$), can be assumed to be from the pristine target. At higher beam energies the current increases and the increasing number densities of Cs ion are embedded below the target surface.

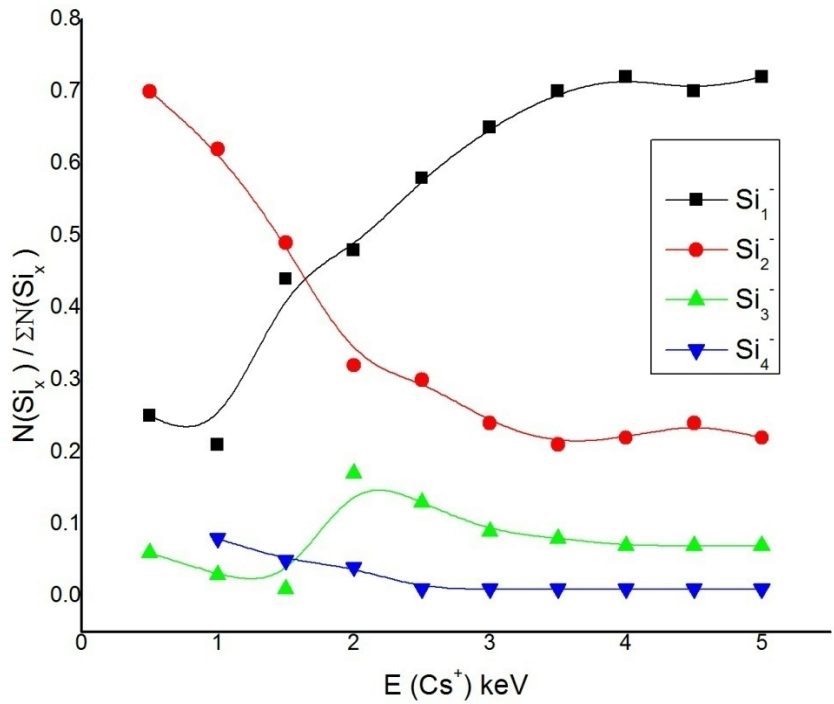

Fig. 2 shows the normalized yields of the four sputtered Si anions from monatomic $Si_1^-$ to $Si_4^-$. The figure provides the clear evidence for the variation of the relative yields as a function of $Cs^+$ energy. The diatomic $Si_2$ yield is highest at 0.5 keV. The cross-over between diatomic and monatomic yields occur at $E(Cs^+) = 2$ keV.

At low $Cs^+$ energies, the emission of a diatomic Si is up to three times more probable as the monatomic Si. The ratio at 0.5 keV reverses at 5.0 keV as can be seen in Fig. 2. The higher clusters $Si_3$ and $Si_4$ have lower relative yields that are consistent and constant as the ionic energy varies within the errors of the experiment. The significant feature of Si sputtering with $Cs^+$ is the mass spectra dominated by the emission of diatomic $Si_2$ at lower energies.

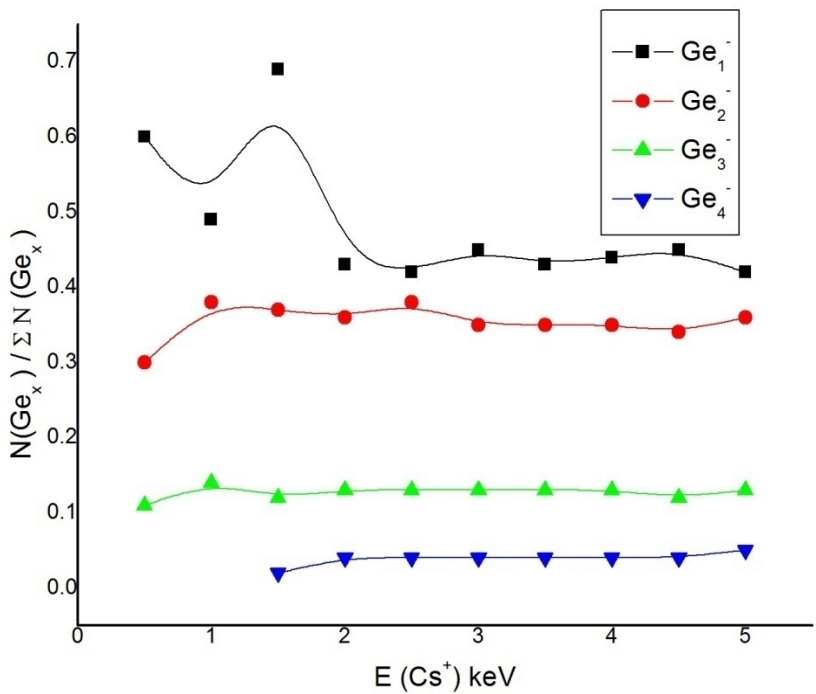

Fig. 3 has the four species sputtered from Ge as a function of E(Cs$^+$). With the exception of the hump in Ge$_1^-$ at E(Cs$^+$)~1.5, all other Ge$_x^-$ show stable normalized yields for the entire energy range. The yields at 0.5 keV can be taken as those from the pristine sample and being more representative of the underlying thermal origin of the emission process.

Fig. 3 shows the normalized anionic yields of the sputtered clusters and Ge atoms as a function of E(Cs$^+$). The experimental data points are connected by curves to guide the eye. The yields follow the consistent pattern for the normalized yield of anions as $N(Ge_1^-) > N(Ge_2^-) > N(Ge_3^-) > N(Ge_4^-)$ for the entire Cs$^+$ energy range. The monatomic $Ge_1^-$ anions have twice the number density of the diatomic ones and six times than that of the tri-atomic clusters. As opposed to the mechanisms of vacancy generation upon energy deposition by Cs$^+$ ions in Si, the Ge matrix produced more uniform energy deposition mechanisms resulting in a consistent and uniform pattern of cluster sputtering.

A typical example of the experimental yields of the sputtered species of ZnO are shown in Fig. 4(a) for the energy of Cs$^+$ = 1.5 keV. This spectrum is chosen to highlight the anomalously high O$^-$ yield as compared to the relatively diminished Zn$^-$ yield. The figure shows

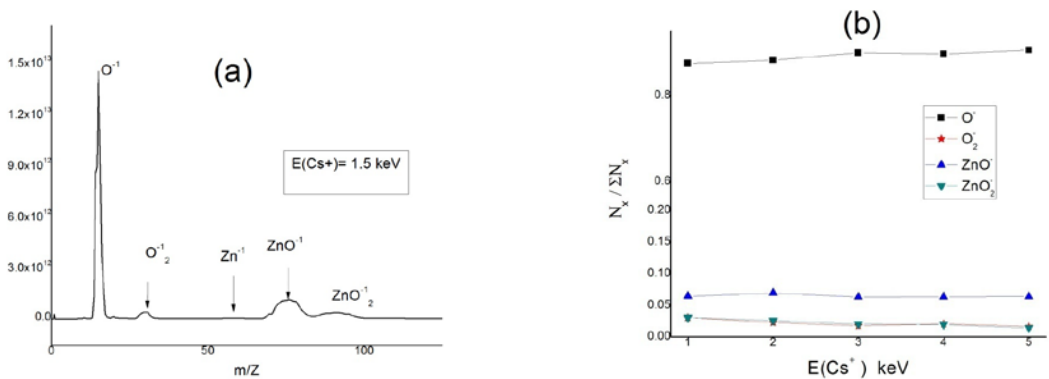

Fig. 4(a) shows the sputtering yields of the four sputtered species from ZnO at E(Cs$^+$) = 1.5 keV and 4(b) has the normalized yield data for the 4 species as a function of Cs$^+$ energy. In Fig 4(b) the atomic O$^-$ has the highest intensity in all the spectra, while Zn$^-$ has 2 orders of magnitude less intensity and is virtually invisible. Molecular O$_2^-$, ZnO$^-$ and ZnO$_2^-$ have the next highest normalized density of anions after O$^-$.

The normalized yields shown in Fig. 4(b) are from the mass spectra of the sputtered species for the range of Cs$^+$ energies E(Cs$^+$) from 1.0 to 5.0 keV. The species with higher intensities are O$^-$ and ZnO$^-$ with O$_2^-$ and Zn$^-$ as the minor species. The SNICS output are anions—a fact that explains the relatively small number densities of Zn$^-$. ZnO solid dissociates as

$$ZnO_s \rightarrow ZnO_g \rightarrow Zn_g^{++} + O_g^{--} \qquad (1)$$

This reaction implies the need for multiple electron capture processes to take place before Zn can escape the surface as an anion. While O$^-$ will naturally emerge as a negative ion. That may explain the much higher number of O$^-$ anions sputtered as compared with the diminished intensities of Zn$^-$ as shown in mass spectra of Fig. 4. The normalized yields $N_x/\Sigma N_x$ are plotted in Fig. 4b as a function of E(Cs$^+$) where $x$ stands for O$^-$, Zn$^-$, O$_2^-$ and ZnO$^-$. The normalized yields of all the sputtered species in the figure are insensitive to the variations in the energy of the irradiating Cs$^+$. This aspect of the experimental results is essential to the understanding and interpretation of the atomic and molecular emissions from ZnO solid. The collision cascades have a direct dependence on irradiating ion energy, which has not been the case in the figures shown above.

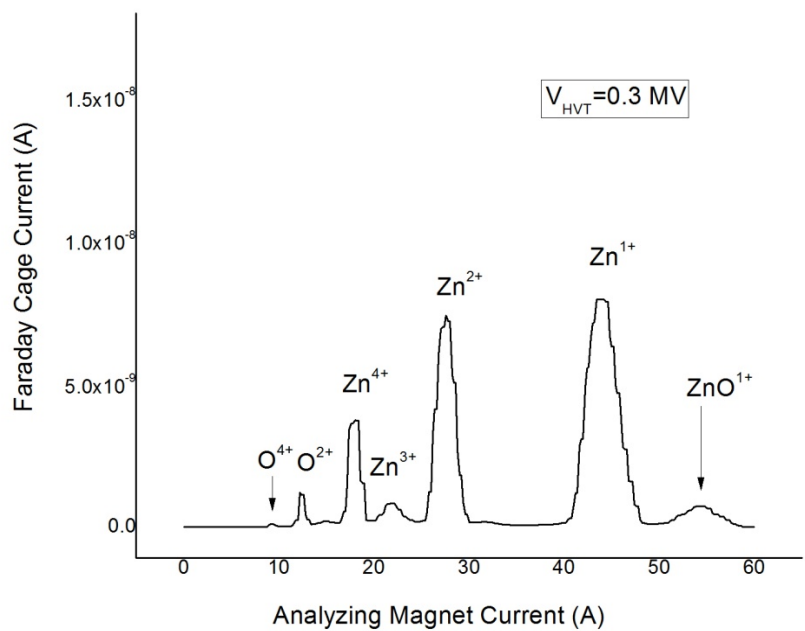

Fig. 5 The mass spectra of multiply charged Zn$^{n+}$ and O$^{n+}$ are shown at 0.3 MeV high voltage terminal (HVT) potential. Multiply charged fragments of ZnO can be seen and the positively charged ZnO$^+$ survives with much reduced intensity.

### 3.2 ZnO$^-$ in the High Voltage Terminal (HVT) of 2 MV Pelletron

A series of experiments was conducted to compare and contrast the Cs$^+$-induced dissociations in SNICS with the dissociation of ZnO$^-$ anions in high energy glancing collisions with the rarified N$_2$ gas molecules in the high voltage terminal-HVT of the 2 MV Pelletron. The aim of the comparison is to visualize the two dissociation mechanisms of the same molecule; one in the condensed phase and the other as a gas molecule. The former is a thermodynamic process and the latter a non-equilibrium, high energy collision mechanism of molecular dissociation and electron removal.

The Pelletron utilizes the technique of the conversion of the accelerated negatively ions into positive ones by charge exchange in HVT. Negative ions extracted from SNICS are sent into the HVT at the desired high voltage. Our objective here is to enumerate the reasons of the relative absence or the much decreased yield of the anion $Zn^-$ in the $Cs^+$-induced dissociations in SNICS ion source, as shown in Fig. 4.

Fig. 5 shows the mass spectra of the cations resulting from the dissociation and multiple-ionization of Zn and O. The multiply charged positive $Zn^{n+}$ (n≥1) ions are orders of magnitude more intense than the corresponding cations of O. The dissociative pattern of ZnO as illustrated in eq. (1) produces positively charged Zn ions in the stripping canal of the HVT. The negative O ions have to lose their negatively bound charge in the stripping canal and charged positively to be extracted out. The multiple electron loss process for O in the spectrum of the cations emerging from HVT as analogous to that in the case of Zn to be extracted as anion in the spectra of anions sputtered from ZnO surface in SNICS.

## 4. Thermal spike induced sublimation

The statistical thermal model was developed to provide justification for cluster emissions from single-walled carbon nanotubes [4]. The model is based on the conjecture that at lower $Cs^+$ energies (≤ few keV), thermal spike may develop where the local temperature $T_s$ is high enough for the sublimation of the constituents of the surface of Si, Ge and ZnO, atoms as well as the clusters. The model was used to describe the cluster emission from mono-shelled nanotubes; it is being extended to describe the emission of clusters from elemental and compound semiconductors.

**4.1 The probability and thermal nature of cluster sputtering**

The probability of creation of a vacancy of *x*-atoms on the surface of the irradiated solid has been evaluated as $\frac{n}{N} = \{(\exp(E_x/kT_s) + 1\}^{-1} = p_x$ , where *n* is the number of vacancies with *x*-atoms that are created in a target region with N atoms. $E_x$ is the respective binding energy of monatomic or the clusters bonded to the matrix of surrounding atoms [4]. Experimentally determined values of the yields of the sputtered clusters $N_x (x \geq 1)$ are obtained from the mass spectra of clusters as a function of $Cs^+$ energy. The normalized yields are $n_x = N_x/\Sigma N_x$. Since it is the experimentally measured density of emission of Si or Ge sputtered species represented as $N_x$, therefore it is directly proportional to the probability of thermally created vacancies $p_x$. The probabilities of emission of any of the cluster species, for example $N_2$, $N_3$, $N_4$ and higher ones, $p_x = n_x/N_S$, where $N_S$ is the total number of atoms in the spike region. $N_S$ is an unknown quantity therefore, we utilize the ratio of the probabilities to eliminate $N_S$. This ratio will allow the calculation of the energies of formation of respective vacancies $E_{xv}$ and the spike temperature $T_S$. These ratios are for the monatomic and diatomic species

$$p_x/p_y = \frac{n_x}{n_y} = \{exp(E_2/kT_S) + 1\}/\{exp(E_1/kT_S) + 1\} \qquad (2)$$

Since in the case of Si and Ge, the monatomic surface binding energies and the sublimation temperature can be obtained from [33,34], we can estimate the energies of formation for the clusters like $Si_2$, $Si_3$ and $Ge_2$ and $Ge_3$. Similar treatment can be extended to the higher clusters provided these are emitted with measurable intensities and at low $Cs^+$ energies, a point that we elaborate here.

It was pointed out in the description of the source SNICS in section 2 that at low $Cs^+$ energy i.e., <1 keV, the beam intensity is low and it spreads on the entire surface of the target. This has two consequences; one is that the Cs penetration is less and secondly, the low intensity beam introduces less damage and one can treat the low energy mass spectra as more representative of the surface effects than the bulk. Therefore, we consider the normalized yields from the mass spectra plotted in Figs. 1 and 2 at $E(Cs^+) = 0.5$ keV for calculations of the energies of formation of di- and tri-atomic Si and Ge.

In the case of normalized yields from the mass spectra of Si in Fig. 1, we can see that the diatomic Si is about three times as that of the monatomic Si. Using eq. (2) with the value of surface binding energy of Si atoms as 4.7 eV [34], and using sublimation temperature as 1908 K for the thermodynamic tables when Si vapour pressure is 1 Pa [33], we obtain $E_v(Si_2) = 4.43$ eV and $E_v(Si_3) = 5.12$ eV. These are the values of the energies of diatomic and tri-atomic vacancy formation on the surface of the Si target. Similar values for Ge are $E_v(Ge_2) = 3.98$ eV and $E_v(Ge_3) = 4.36$ eV. While the value for single atom removal from Ge surface is 3.88 eV [34].

The above analysis of the normalized sputtering yields of clusters from Cs-irradiated Si and Ge assumes the initiation of thermal spikes on and around the surface. The inputs to the model are the ratios of the normalized yields, the sublimation temperature and the energy of formation of at least one of the sputtered constituents.

**3.4 Rates of ZnO dissociation**

The case of sputtering of ZnO has two dimensions. Its irradiation as a solid with $Cs^+$ has energy deposition process with similar collision cascades that occur in Si or Ge, however, as a molecular solid, the constitution of the cascades is different. Zn and O with different masses receive different amounts of energies in collision, thereafter the twin collisions progress simultaneously but differently. Monte Carlo simulations like SRIM [34] may yield better insight into the spreading and merging of the twin cascades but we are interested here in the net effect of the irradiation i.e., the thermal spike generation that raises the surface temperatures and the material sublimes. The thermal model has a justification for application because the simulations neither consider thermal spikes nor provide justification for the emission of clusters.

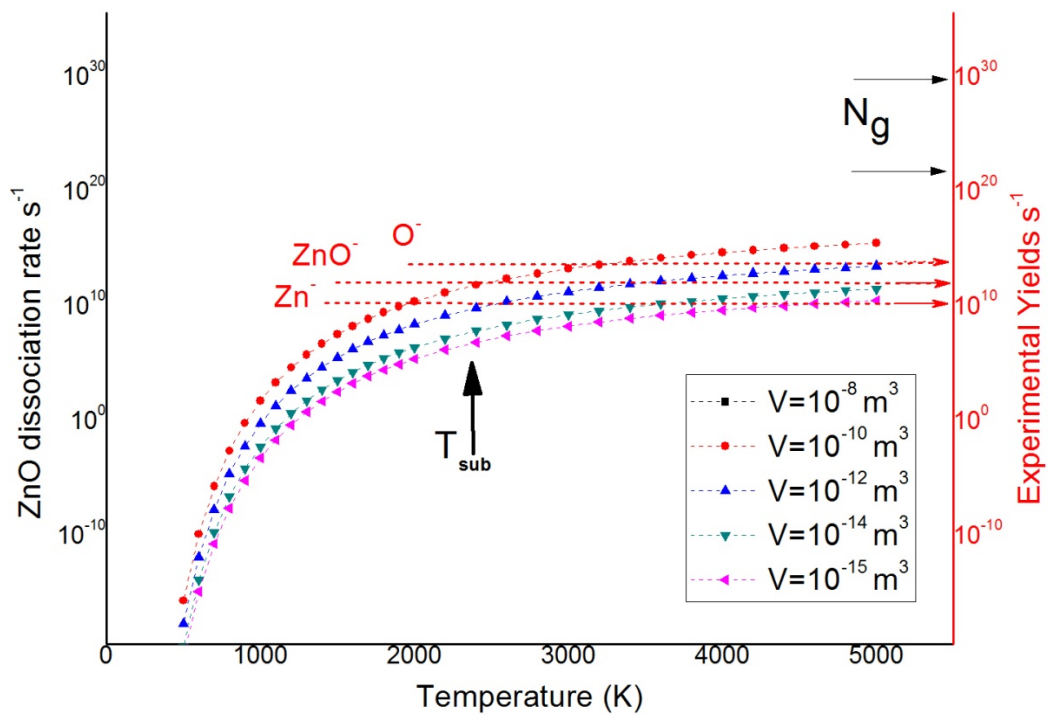

Fig. 6. The calculated rates of dissociation of ZnO and the experimental yields of $ZnO^-$ are plotted for a range of volume V, as a function of temperature. Experimental yields of $O^-$ and $Zn^-$ are also indicated. The range of gaseous ZnO from a gas-solid mixture in thermodynamic equilibrium $N_g$, are indicated by arrows.

We consider the route illustrated in eq. (1) as the process of energy deposition in ZnO and calculate the rates of dissociation after the solid has sublimed into gaseous species. In Fig. 6 the rates of the dissociation reaction of ZnO from eq. (1) are plotted as a function of the local spike temperature. The rate equation for the dissociation of ZnO to Zn and O is

$$K = \{[Q(Zn)Q(O)/Q(ZnO)]\exp(-\frac{E_{dis}}{kT_{sub}})\} \qquad (3)$$

where Qs are the respective partition functions per unit volume for the molecule ZnO and its atomic constituents, $E_{dis}$ is the dissociation energy and $T_{sub}$ is the sublimation temperature of ZnO. The value of $E_{dis}$ is taken as 3.5 eV and $T_{sub}$ is expected to be around 2800 K [35]; the calculations show that to be the case. The volume V for the gaseous species has been varied between the limits of a monolayer equivalent to V=$10^{-15}$ m$^3$ to the entire volume of SNICS source V=$10^{-8}$ m$^3$.

In Fig. 6, in the temperature ranges from 500 to 5000 K, for the minimum volume V, the ZnO dissociation rates can be seen to vary between $10^{-20}$ to ~$10^2$ s$^{-1}$. For T >1500 K, the calculated dissociation rates gradually approach the experimentally observed yields of O$^-$ and ZnO$^-$. With V between $10^{-15}$ and $10^{-12}$ m$^3$, one gets a quantitative agreement between the calculated dissociation rates and the experimentally observed sputtering yields of the atomic and molecular species.

We must make a clear distinction between the localized thermal spike and the whole solid subliming by calculating the number densities required to achieve an overall equilibrium between solid ZnO$_s$ and it gaseous counterpart ZnO$_g$. The partition function of the gas and solid $Q_g$ and $Q_s$ can be computed from individual gas molecule ($q_g$) and oscillator ($q_s$) [36]. For the equilibrium condition between the whole solid in equilibrium with its gaseous components is

$$N_g = q_g(T,V)/q_s(T) \qquad (4)$$

This value is indicated for the two extreme values of the volume by two arrows in the top right hand corner of Fig. 6. These values of $N_g$ are higher by four to ten orders of magnitude as compared with the experimentally observed number densities of the sputtered species and those calculated from the dissociation rate equation (3) by assuming localized thermal spikes. Therefore, we conclude that ion-induced thermal spikes have a local character that should not be confused with the sublimation of the whole solid.

## 5. Space-filling, multifractal description of cascades-to-spike transition

The transition from binary collision cascades to thermal spikes can be described by treating the tree of binary cascades as fractal whose dimension increases as the energy of recoils reduces. Following Mandelbrot [1] collisions induced in a plane of atoms follow a tree-like branching effect. The same angle between the branches occurs at each branching point. The lengths of successive branches decrease depending upon the nature of the interatomic potential V(r). The ratio of the lengths between successive branches, $\gamma$, remains fixed. The fractal dimension for our binary-branching collision cascade is

$$D = \ln 2 / \ln(1/\gamma) \tag{5}$$

A collision cascade is governed by the inverse-power potential of the form

$V(r) \propto r^{-1/m}$ where $0 \leq m \leq 1$ [7,37]. Consider a cascade started with initial energy $E_0$. For a binary cascade the mean energy for the recoiling partners is $\bar{E}_1 = \frac{1}{2} E_0$. For the nth generation of collisions it will be $\bar{E}_n = \frac{1}{2} \bar{E}_{n-1}$. The ratio of the successive mean free paths is $\gamma = \lambda_n/\lambda_{n-1}$. The fractal dimension D =1 for the first branching at $\gamma = \frac{\lambda_n}{\lambda_{n-1}} = \frac{1}{2}$, the potential $V(r) \propto r^{-2}$. Similarly D =2 for $\gamma = (\frac{1}{2})^{1/2}$ with $V(r) \propto r^{-4}$. Thus the branching ratio $\gamma = 2^{-2m}$ for $V(r) = G r^{-1/m}$ with $0 \leq m \leq 1$. This yields the fractal dimension of a collision cascade as a function of the exponent $m$ in the form

$$D = \frac{\ln 2}{\ln(1/\gamma)} = \frac{1}{2m} \tag{6}$$

Effectively, the fractal dimension has been related to the changes of the effective interatomic potential. These relations show that the cascades' fractal dimension is inversely proportional to the exponent of the interatomic potential. Consequent to the transition of potential V(r) from m=1/2 to 1/4 to 1/6, the fractal dimensions vary from 1 to 2 to 3. When the fractal dimension D and the physical dimension of the solid are same, the space-filling fractal emerge. For this case the collision mean free paths are of the same order as the interatomic spacing which sets the majority of atoms in motion. This happens in a limited volume where the equipartition of potential and kinetic energies occurs that lead cascades to approach localized thermal equilibrium.

## 6. Conclusions

We conclude that the sputtering of atomic and molecular species from heavy ion-irradiated elemental and molecular solids occurs due to the transition of the binary collision cascade to thermal spike. This transition occurs locally for each irradiating ion. The self-similarity of the branching tree of binary collisions provides the multifractal description of this transition that culminates as the space-filling thermal spikes. The space-filling fractal has a local character; hence the thermal spikes are localized around the track of the expanding, energy sharing cascade among the increasing number of neighbors. We are proposing that each irradiating $Cs^+$ ion in Si, Ge and ZnO solid ends in a localized thermal spike whose proximity to the outer surface determines the rates of the subliming and sputtering atoms and molecules. The proposed model for the localized thermal spike as the space-filling multifractal may also explain the emission of clusters from $sp^2$-bonded solids like graphite and carbon allotropes fullerenes and nanotubes where clusters are sputtered species with much higher number densities compared with the atomic carbon [2-4].

**ACKNOWLEDGEMENTS**

The experimental work was performed at the 2 MV Pelletron Laboratory at CASP, Government College University (GCU), Lahore, Pakistan. Technical help provided by Mr. M. Khalil is gratefully acknowledged.

**Contributions of Authors:** The first author (S.A.) supervised the experiment and developed space-filling fractal model, all other authors contributed equally to the execution of the experiments and data management.